\documentclass[aps,prf,twocolumn,groupedaddress,amsmath,amssymb,longbibliography,nofootinbib]{revtex4-1}
\usepackage{graphicx}  
\usepackage{dcolumn}   
\usepackage{bm}        
\usepackage{verbatim}   
\usepackage{xcolor}   

\usepackage{braket}
\usepackage{pstricks}
\usepackage{epsfig}

\newcommand*{\Ta}{{\rm Ta}}

\newcommand{\Ree}{{\rm Re}}
\newcommand{\Ros}{{\rm Ro}}
\newcommand*{\Pec}{{\rm Pe}}
\newcommand*{\Nus}{{\rm Nu}}
\newcommand*{\Ray}{{\rm Ra}}

\newcommand{\Pra}{{\rm Pr}}

\newcommand{\kv}{\ensuremath{\mathbf{k}}}
\newcommand{\rv}{\ensuremath{\mathbf{r}}}
\newcommand{\ve}{{\mathbf{v}}}
\newcommand{\lp}{\ensuremath{\left(}}
\newcommand{\rp}{\ensuremath{\right)}}

\begin{document}

\title{Polar waves and chaotic flows in thin rotating spherical
  shells}

\author{F. Garcia}
\affiliation{Department of Magnetohydrodynamics, Helmholtz-Zentrum Dresden-Rossendorf, Bautzner Landstra\ss e 400, D-01328 Dresden, Germany}
\affiliation{
Anton Pannekoek Institute for Astronomy, University of Amsterdam, Postbus 94249, 1090 GE Amsterdam, The Netherlands  
}

\author{F.~R.~N. Chambers}
\affiliation{
Anton Pannekoek Institute for Astronomy, University of Amsterdam, Postbus 94249, 1090 GE Amsterdam, The Netherlands    
}
\author{A.~L. Watts}
\affiliation{
Anton Pannekoek Institute for Astronomy, University of Amsterdam, Postbus 94249, 1090 GE Amsterdam, The Netherlands    
}

\date{\today}

\begin{abstract}
Convection in rotating spherical geometries is an important physical
process in planetary and stellar systems. Using continuation methods
at low Prandtl number, we find both strong equatorially asymmetric and
symmetric polar nonlinear rotating waves in a model of thermal
convection in thin rotating spherical shells with stress-free boundary
conditions.  For the symmetric waves convection is confined to high
latitude in both hemispheres but is only restricted to one hemisphere
close to the pole in the case of asymmetric waves. This is in contrast
to what is previously known from studies in the field. These periodic
flows, in which the pattern is rotating steadily in the azimuthal
direction, develop a strong axisymmetric component very close to
onset.  Using stability analysis of periodic orbits the regions of
stability are determined and the topology of the stable/unstable
oscillatory flows bifurcated from the branches of rotating waves is
described. By means of direct numerical simulations of these
oscillatory chaotic flows, we show that these three-dimensional
convective polar flows exhibit characteristics, such as force balance
or mean physical properties, which are similar to flows occuring in
planetary atmospheres. We show that these results may open a route to
understanding unexplained features of gas giant atmospheres, in
particular for the case of Jupiter.  These include the observed
equatorial asymmetry with a pronounced decrease at the equator (the
so-called dimple), and the coherent vortices surrounding the poles
recently observed by the Juno mission.
\end{abstract}

\maketitle

\section{Introduction}

The problem of thermal convection in rotating spherical geometries is
of central importance in planetary science and astrophysics.
Planetary dynamos~\cite{DoSo07}, zonal jets in giant planet
atmospheres~\cite{JoKu09,CAFL17}, differential rotation of layers in
solar-like stars~\cite{MMK15} and convection driven by nuclear
reactions in the oceans of neutron stars~\cite{Wat12}, for example,
all share the key ingredients of temperature gradients, rotation and
spherical geometry.  A widely used model that accounts for these
factors is the Boussinesq approximation of the Navier-Stokes and
energy equations in a rotating frame of reference~\cite{SiBu03}. The
problem is then described by the aspect ratio $\eta=r_i/r_o$, where
$r_i$ ($r_o$) is the radius of the inner (outer) spherical boundary,
the Prandtl, $\Pra$, and Taylor, $\Ta$, numbers characterise the
relative importance of viscous (momentum) diffusivity to thermal
diffusivity, and rotational to viscous forces, respectively, and the
Rayleigh number $\Ray$ is associated with buoyancy forces.

For small temperature differences between the boundaries (small
$\Ray$), heat is conducted to the outer boundary and the fluid is at
rest. At a critical Rayleigh number $\Ray_c$ convection sets in, and
its preferred mode pattern depends strongly on $\eta$, $\Pr$ and
$\Ta$. In the case of the rapidly rotating ($\Ta>10^{10}$) thin shells
($\eta=0.9$), which may be appropriate for modelling Jupiter's
atmosphere~\cite{HAW05} or stellar convective oceans~\cite{GCW18}, the
convective pattern strongly depends on $\Pr$: spiralling columnar
modes~\cite{Zha92} in the bulk of the fluid for large $\Pr>0.1$;
equatorial inertial modes~\cite{Zha93} lying close to the outer sphere
and at low latitudes for moderate $10^{-3}<\Pr<10^{-2}$; and polar
modes~\cite{GSN08,GCW18} confined to large latitudes for both small
$\Pr<10^{-3}$ and moderate $10^{-2}<\Pr<10^{-1}$).

Inertial modes, characteristic of small Prandtl number $\Pr<1$, are
explained in terms of solutions of the inviscid problem described by
the Poincar\'e equation~\cite{Zha93,Zha94,BuSi04}. By increasing the
size of the inner core~\cite{Zha93} showed that equatorial inertial
modes tend to be located at higher latitudes, especially for low wave
numbers $m$. For sufficiently large radius ratio these modes can be
interpreted as polar modes, as convection is restricted to high
latitudes. With small but nonzero viscosity (i.~e. the small $\Pr$
number thermal convection problem) the situation is similar and
inertial modes located at high latitudes (polar modes) seem to be
favoured (instead of equatorial modes) when the radius ratio is
large~\cite{GSN08,GCW18}. Indeed, as inertial modes of the Poincar\'e
equation the onset of convection at small $\Pr$ can be antisymmetric
with respect to the equator~\cite{GSN08,GCW18}.

The large and moderate $\Pr$ number nonlinear regimes have been
studied extensively ~\cite{Zha92,SiBu03,HAW05} but low-$\Pr$ fluids of
most relevance to planetary and stellar systems have received less
attention. Very recently, numerical (\cite{HoSc17} for rotating plane
layer or ~\cite{KSVC17,LKZ18} for full sphere), and
laboratory~\cite{ABGHV18} (cylindrical container) studies have
revealed immense complexity. Low $\Pr$ convection can be strongly
oscillatory even right at the onset, involving several modes,
and without the appearance of the steady drifting waves that are
characteristic of large $\Pr$. Convection can even be
sub-critical~\cite{KSVC17} at sufficiently large $\Ta$. The latter
studies are devoted to the study of planetary cores involving nonslip
boundaries, and in the case of a full rotating sphere~\cite{KSVC17}
convection sets in via quasigeostrophic spiralling or equatorial
modes, both being equatorially symmetric, as happens in most of the
studies in rotating spherical geometry.

Low-$\Pr$ number nonlinear flows in a regime where convective onset
occurs via polar equatorially antisymmetric modes have not been
studied to date and this is the main purpose of the present
study. These modes are characteristic of thin shells, and have
application to both stellar oceans (which may have very low
$\Pr=10^{-6}$~\cite{GCW18}) and Jupiter's atmosphere, where they may
be the preferred form of onset~\cite{GCW18}. Equatorial symmetry in
three-dimensional deep convection models (which allow strong zonal
jets to extend from the surface towards the planet's interior
contrasting to the shallow layer approach) is a very topical issue,
because the antisymmetric component of the flows is directly related
to the odd gravity harmonics recently measured in~\cite{Kas_et_al18}
and used to infer the interior structure of Jupiter's atmosphere. Its
strong positive jet, at around $25^\circ$ latitude, dominates the
antisymmetric component of the measured
flow~\cite{Kas_et_al18}. Current deep convection models of
Jupiter~\cite{HAW05} can reproduce the strong positive equatorial jet
velocities and its latitudinal extend. By incorporating the effect of
radiative heating of the atmosphere and magnetic dissipation deep in
the atmosphere at high latitudes, ~\cite{ScLi09} demonstrated that
equatorial Rossby waves relate to positive equatorial jets, and that
radiative effects are responsible for jets at higher latitudes. The
latter studies (\cite{HAW05,ScLi09}) however, reproduce neither the
observed Jovian equatorial asymmetry~\cite{Kas_et_al18} (traditionally
associated with the Great Red Spot~\cite{ReHi83}, see~\cite{Mar93} for
a review) nor the pronounced dimple. The latter was discussed
in~\cite{GWA13} in the context of stratified anelastic models. The
anticyclonic coherent vortices observed in the giant planet's jets
have been only reproduced when considering a combined shallow-water
deep non-magnetic approach~\cite{HGW15}, but again lack equatorial
asymmetry. The very recent experiments of~\cite{CAFL17} provide strong
support for deep rather than shallow-water modelling. High latitude
jets were obtained even when considering viscous dissipation,
mimicking the expected braking of the jets due to Lorentz
forces~\cite{CAFL17}.  Because the polar modes studied in this paper
naturally develop high latitude jets, and because the onset of
convection may be of polar type when considering nonslip
boundaries~\cite{GSN08}, our results are consistent with the
experiments, and provide evidence that equatorial asymmetry and high
latitude jets and vortices could develop in deep convection models of
spherical geometry.

We use continuation techniques~\cite{Kel77,SaNe16} to obtain nonlinear
periodic flows (rotating waves) bifurcated from the conductive
motionless state and we study their stability. This allows us to
describe the patterns of the perturbations giving rise to oscillatory
flows which are obtained by means of direct numerical simulations
(DNS).  We adopt $\eta=0.9$, $\Pra=3\times 10^{-3}$ and $\Ta=10^7$,
with stress-free boundary conditions, and consider $\Ray$ the control
parameter for our study. We emphasise that in this regime polar modes
are linearly preferred~\cite{GCW18} and the nonlinear saturation of
this recently discovered instability~\cite{GSN08,GCW18} is still
unknown. Polar modes are preferred as well for $\Ta>10^{10}$ so the
present study indicates the need for further research at larger $\Ta$
which is the relevant regime for Jovian atmosphere dynamics. The paper
is organized as follows. In \S\ \ref{sec:model} we introduce the
formulation of the problem, and the numerical method used to obtain
the solutions. A brief description of the continuation method and the
stability analysis of the rotating waves is provided as well. In
\S\ \ref{sec:res} the results are presented: the bifurcation diagrams
and the patterns of the waves and their eigenfunctions are described,
the study of chaotic flows is undertaken, and the application to gas
giant planetary atmospheres and other physical context of the results
is discussed (tentatively). Finally \S\ \ref{sec:sum} summarises the
results obtained.

\section{Model}
\label{sec:model}

Boussinesq thermal convection in a rotating spherical shell is
considered. The fluid is homogeneous with density $\rho$ and constant
physical properties: thermal diffusivity $\kappa$, thermal expansion
coefficient $\alpha$, and dynamic viscosity $\mu$. The shell is
defined by its inner and outer radius $r_i$ and $r_o$, and it is
rotating with constant angular velocity ${\bf \Omega}=\Omega {\kv}$
about the vertical axis. A radial gravitational field ${\bf g}=-\gamma
\rv$ ($\gamma$ is constant and $\rv$ the position vector) is imposed
and $\rho=\rho_0(1-\alpha(T-T_0))$ is assumed in just the
gravitational term. In the other terms a reference state
$(\rho_0,T_0)$ is assumed (see for instance~\cite{Ped79,Cha81}).

On the perfectly conducting boundaries a temperature difference is
imposed $\Delta T=T_i-T_o$, $T(r_i)=T_i$ and $T(r_o)=T_o$ and
stress-free boundary conditions are used for the velocity
field. Stress-free conditions are appropriate for the study of
planetary atmospheres~\cite{HAW05} as well as stellar convective
zones~\cite{GCW18}. The mass, momentum and energy equations are
derived in the rotating frame of reference as in~\cite{SiBu03}. This
frame of reference rotates westward, following the planetary rotation.
The equations are expressed in terms of velocity ($\ve$) and
temperature ($\Theta=T-T_c$) perturbations of the basic conductive
state $\ve=0$ and $T_c(r)=T_0+\eta d \Delta T(1-\eta)^{-2}r^{-1}$,
$\eta=r_i/r_o$ being the aspect ratio, $d=r_o-r_i$ being the gap
width, and $T_0=T_i-\Delta T(1-\eta)^{-1}$ being a reference
temperature.  With units $d$ for the distance, $\nu^2/\gamma\alpha
d^4$ for the temperature, and $d^2/\nu$ for the time, the equations
are
\begin{align}
&\nabla\cdot\ve=0,\label{eq:cont}\\
&\partial_t\ve+\ve\cdot\nabla\ve+2\Ta^{1/2}\kv\times\ve = 
-\nabla p^*+\nabla^2\ve+\Theta\rv,\label{eq:mom}\\
&\Pr\lp\partial_t \Theta+\ve\cdot\nabla \Theta\rp= \nabla^2
\Theta+\Ray \eta (1-\eta)^{-2}r^{-3} \rv\cdot\ve,  \label{eq:ener}
\end{align}
where $p^*$ is a dimensionless scalar containing all the potential
forces. We neglect centrifugal effects by assuming $\Omega^2/\gamma
\ll 1$, as is usual for geophysical and astrophysical
applications. With the above considerations, four non-dimensional
parameters -the aspect ratio $\eta$ and the Rayleigh $\Ray$, Prandtl
$\Pr$, and Taylor $\Ta$ numbers- describe the physics of the
problem. These numbers are defined by
\begin{equation*}
  \eta=\frac{r_i}{r_o},\quad
  \Ray=\frac{\gamma\alpha\Delta T d^4}{\kappa\nu},\quad
  \Ta^{1/2}=\frac{\Omega d^2}{\nu},\quad
  \Pr=\frac{\nu}{\kappa}.
\end{equation*}

To solve the model equations~\ref{eq:cont}-\ref{eq:ener} with the
prescribed boundary conditions a pseudo-spectral method is used
(see~\cite{GNGS10} and references therein). In the radial direction a
collocation method on a Gauss--Lobatto mesh is considered whereas
spherical harmonics are used in the angular coordinates. The
incompressibility condition leads to the so-called toroidal/poloidal
decomposition for the velocity field~\cite{Cha81}. The code is
parallelized in the spectral and in the physical space using OpenMP
directives. We use optimized libraries (FFTW3~\cite{FrJo05}) for the
fast Fourier transforms FFTs in longitude and matrix-matrix products
(dgemm GOTO~\cite{GoGe08}) for the Legendre transforms in latitude
when computing the nonlinear terms.

High order implicit-explicit backward differentiation formulas
(IMEX--BDF)~\cite{GNGS10,GNS13} are used for time-stepping the
discretized equations. In the IMEX method we treat the nonlinear terms
explicitly, in order to avoid solving nonlinear equations at each time
step. The Coriolis term is treated fully implicitly to allow larger
time steps. The use of \textit{matrix-free} Krylov methods (GMRES in
our case) for the linear systems facilitates the implementation of a
suitable order and time stepsize control. An accurate efficient
time-stepper is necessary for successfully applying the continuation
method due to the high resolutions demanded for the present study.

\section{Continuation method and stability analysis for rotating waves}
\label{sec:co_meth}

The study of the patterns of rotating waves (RW) and the analysis
  of their stability is important because it characterises the
  symmetry~\cite{CrKn91} of the oscillatory solutions (modulated
  rotating waves~\cite{Ran82,CoMa92}) bifurcated from the branches
  (see also~\cite{GLM00}). Indeed, the interaction of solutions of
  different symmetry class (for instance equatorial symmetry in low
  order models of rotating systems~\cite{KnLa96}) gives rise to
  complex dynamics.  These secondary oscillatory solutions might play
  a key role in organizing the global dynamics~\cite{KUL12} and thus
  chaotic flows close to the onset characteristic of low-$\Pr$
  fluids~\cite{HoSc17}. The relevance of unstable RW for the
  understanding of turbulent flows~\cite{Hof_et_al04} makes their
  study of importance. Continuation methods are powerful as they allow
  the tracking of curves of unstable RW which can not be obtained by
  means of DNS. We note that for the study of periodic, quasiperiodic
  and chaotic flows (typical in thermal rotating systems~\cite{HLR94})
  mode decomposition techniques~\cite{WSK13,LeVe17} based on DNS are a
  powerful tool as well, as they allow the identification of the
  relevant modes and patterns of the flow for very challenging
  problems. However, since the pioneering experiments of R. Hide on
  rotating thermal convection~\cite{Hid53} (see~\cite{GRS10} for a
  review) the scientific community has started to understand the
  origin of these patterns and nature of chaotic geophysical flows
  using bifurcation and dynamical systems
  theory~\cite{RuTa71,CrKn91,Kuz98}. Continuation methods are a basic
  tool for their analysis~\cite{Doe86,DoTu00} and a brief description
  is included in this section. The interested reader is referred
to~\cite{SNGS04b}, or the comprehensive tutorial~\cite{SaNe16}, for a
theoretical description and implementation of this tool. An
application of the method to thermal convection in rotating spherical
shells can be found in~\cite{GNS16}. The continuation techniques used
here rely on time integrations of the Navier-Stokes plus energy
equations (Eqs.~\ref{eq:cont}-\ref{eq:ener}). The discretized system
is of dimension $n=(3L_{\text{max}}^2+6L_{\text{max}}+1)(N_r-1)$,
$L_{\text{max}}$ being the spherical harmonic truncation parameter and
$N_r$ the number of collocation points in the radial direction, and
takes the form
\begin{align}
L_0\partial_tu=L u+B(u,u),
\label{eq:op_eq}
\end{align}
where $u$ contains the spherical harmonic amplitudes of the toroidal,
poloidal and the temperature perturbation scalars at the radial
collocation mesh. Here $L_0$ and $L$ are linear operators which
include the boundary conditions (see~\cite{GNGS10} for details). The
operator $L$ depends on $\Ray$ (the control parameter of the present
study) and includes all the linear terms and the bilinear operator $B$
only contains the non-linear (quadratic) terms.

To study the dependence of an azimuthally rotating wave
(see~\cite{Ran82,GLM00} and references therein for a mathematical
definition and theory of these waves), with frequency $\omega$ and
with $m_d$-fold azimuthal symmetry, on the parameter $p=\Ray$,
pseudo-arclength continuation methods for periodic orbits are
used~\cite{SNGS04b}. They allow to obtain the curve (branch) of
periodic solutions $x(s)=(u(s),\tau(s),p(s))\in\mathbb{R}^{n+2}$, $u$
being the rotating wave, $\tau=2\pi/(m\omega)$ the rotation period,
and $s$ the arclength parameter. The method requires adding the
pseudo-arclength condition
\begin{equation}
h(u,\tau,p)\equiv\langle w,x-x^0 \rangle=0,
\end{equation}
$x^0=(u^0,\tau^0,p^0)$ and $w=(w_u,w_\tau,w_p)$ being the predicted
point and the tangent to the curve of solutions ($\langle .,. \rangle$
stands for the inner product in $\mathbb{R}^{n+2}$), respectively,
obtained by extrapolation of the previous points along the curve.

The system which determines a single solution $x=(u,\tau,p)$ on the
branch is:
\begin{equation}
H(u,\tau,p)= \left(
\begin{array}{c}
u-\phi(\tau,u,p)\\
g(u)\\
h(u,\tau,p)\\
\end{array}
\right)
=0,
\label{eq:H_eq}
\end{equation}
where $\phi(\tau,u,p)$ is a solution of
Eqs.~\ref{eq:cont}-\ref{eq:ener} at time $\tau=2\pi/(m\omega)$ and
initial condition $u$ for fixed $p$. The condition $g(u)=0$ is
  selected to fix the undetermined azimuthal phase of the rotating
  wave with respect to the rotating reference frame.  We use
$g(u)=\langle u,\partial_{\varphi}u_c \rangle$ where $u_c$ is a
reference solution (a previously computed rotating wave, or the
preferred mode at the onset). It is a necessary condition for
$\|u-u_c\|^2_2$ to be minimal with respect to the phase
(see~\cite{SGN13}). For the computation of the inner products
$\langle\cdot,\cdot\rangle$ between two functions expanded in
spherical harmonics, we use the definitions of~\cite{SGN13}.

To solve the large non-linear system defined by Eq.~(\ref{eq:H_eq}) we
use Newton-Krylov methods. These are matrix-free methods that do not
require the explicit computation of the Jacobian
$D_{(u,\tau,p)}H(u,\tau,p)$, but only its action on a given vector
which consists of a time integration of a system (of dimension $2n$)
obtained from the Navier-Stokes and energy equations. For the linear
systems we use GMRES~\cite{SaSc86}. Due to the particular form of the
spectrum of $D_{(u,\tau,p)}H(u,\tau,p)$ for dissipative systems, GMRES
does not need preconditioning~\cite{SNGS04b}. We note that periodic
rotating waves can be also obtained efficiently by Newton-Krylov
continuation methods but as steady solutions of the equations written
in a reference frame which is rotating with the wave, see for
instance~\cite{SGN13} for thermal convection in spherical geometries
or~\cite{TLW19} for pipe flow.

The stability of a periodic solution is determined following Floquet
theory~\cite{JoSm07}.  It requires the computation of the dominant
eigenvalues and eigenfunctions of the map $\delta u\longrightarrow D_u
\phi(\tau,u,p)\delta u= v(\tau)$, with $v(\tau)$ being the solution of
the first variational equation, obtained by integrating the system
\begin{align*}
&\partial_t z=L_0^{-1}(L(p)z+B(z,z)),\\
&\partial_t v=L_0^{-1}(L(p)v+B(z,v)+B(v,z)), 
\end{align*}
of dimension $2n$, with initial conditions $z(0)=u$ and $v(0)=\delta
u$, over a rotation period $\tau$, with fixed $p$.

The ARPACK package is used to obtain the eigenvalues of the map with
larger modulus corresponding to the dominant complex Floquet
multipliers $\lambda=|\lambda|e^{i\text{Arg} \lambda}$. Once the
dominant Floquet multipliers cross the unit circle boundary
($|\lambda|>1$) the rotating wave becomes unstable. The marginal
Floquet multiplier with associated eigenfunction $v_1=\partial_t u$
lying on the unit circle, appearing due to the invariance under
azimuthal rotations, is deflated by computing the eigenvalues of the
map $\delta u \longrightarrow v(\tau)-\braket{v(\tau),v_1}v_1$ to
avoid unnecessary computations.

We note that the method is computationally demanding because it requires
the time integration of an ODE system of dimension $2n$ over one
rotation period, and thus an efficient time-stepper is
mandatory. Because the periodic orbit is a rotating wave there is a
more efficient alternative to this procedure~\cite{SGN13,Tuc15} which
consists of studying the stability as a fixed point of a vector field.

\begin{table*}[t!] 
  \begin{center}
    \begin{tabular}{lcccccccccc}
      \vspace{0.1cm} $m_d$ & $N_r$ & $L_{\text{max}}$ & $L_{\text{max}}/m_d$ & $\Ray$ & $\omega$ & $K$ & $|\lambda|$ & $\text{Arg} \lambda$ & $n$ & $n_d$\\[0.pt]
      \hline\\[-8.pt]
      $19$ &$30$ & $114$ & $6$ & $310$ & $169.57655$ & $6.79415$ & $0.996526$ & $0.423632$ & $1150517$ & $60581$\\
      $19$ &$50$ & $190$ & $10$ & $310$ & $169.57656$ & $6.79407$ & $0.996523$ & $0.423626$ & $5362609$ & $282289$\\ \hline
  \end{tabular}
  \caption{Spatial discretization study. Frequency $\omega$,
    volume-averaged kinetic energy $K$, and dominant complex Floquet
    multiplier $\lambda=|\lambda|e^{i\text{Arg} \lambda}$, versus the
    number of radial collocation points $N_r$ and the spherical
    harmonics truncation parameter $L_{\text{max}}$. The dimension of
    the system is $n=(3L_{\text{max}}^2+6L_{\text{max}}+1)(N_r-1)$ for
    the stability analysis and
    $n_d=((3L_{\text{max}}^2+6L_{\text{max}})/m_d+1)(N_r-1)$ when
    assuming $m_d$-fold azimuthal symmetry to obtain the rotating
    waves.}
  \label{table:mesh_study}
  \end{center}
\end{table*}

\section{Results}
\label{sec:res}

Planetary atmospheres and convective stellar regions have low $\Pra$
and high $\Ta$ and are modelled with thin spherical geometry
$\eta=0.9$ and stress-free boundary conditions.  For a neutron star
ocean, $\Ta>10^{20}$ and $\Pra<10^{-3}$~\cite{GCW18}, while for the
Jovian atmosphere $\Ta>10^{30}$ and $\Pra\lesssim
10^{-1}$~\cite{HAW05}. In addition, thermally driven liquid metal
  cores are not far from this regime but the spherical shell is thick
and nonslip, as is Earth's outer core~\cite{Jon11}. According to the
linear studies~\cite{GSN08,GCW18} equatorially antisymmetric or
symmetric polar modes are good candidates to be linearly dominant in
such regimes, but the nonlinear saturation of these modes has never
been studied up to date. This is the purpose of the present study
which numerically investigates the finite amplitude convection at
$\eta=0.9$, $\Pra=3\times 10^{-3}$ and $\Ta=10^7$. Although the latter
value is still far from real applications, it belongs to the parameter
regime of equatorially antisymmetric polar modes, and thus explores an
exciting regime for astrophysical/planetary convection. The
very large $\eta$ and low $\Pra$ used lead to very challenging small
spatial and time scales and thus a moderate, but still relevant, $\Ta$
is considered to make the problem computationally feasible.

\subsection{Bifurcation diagrams of polar rotating waves}

\begin{figure*}
\includegraphics[scale=1.2]{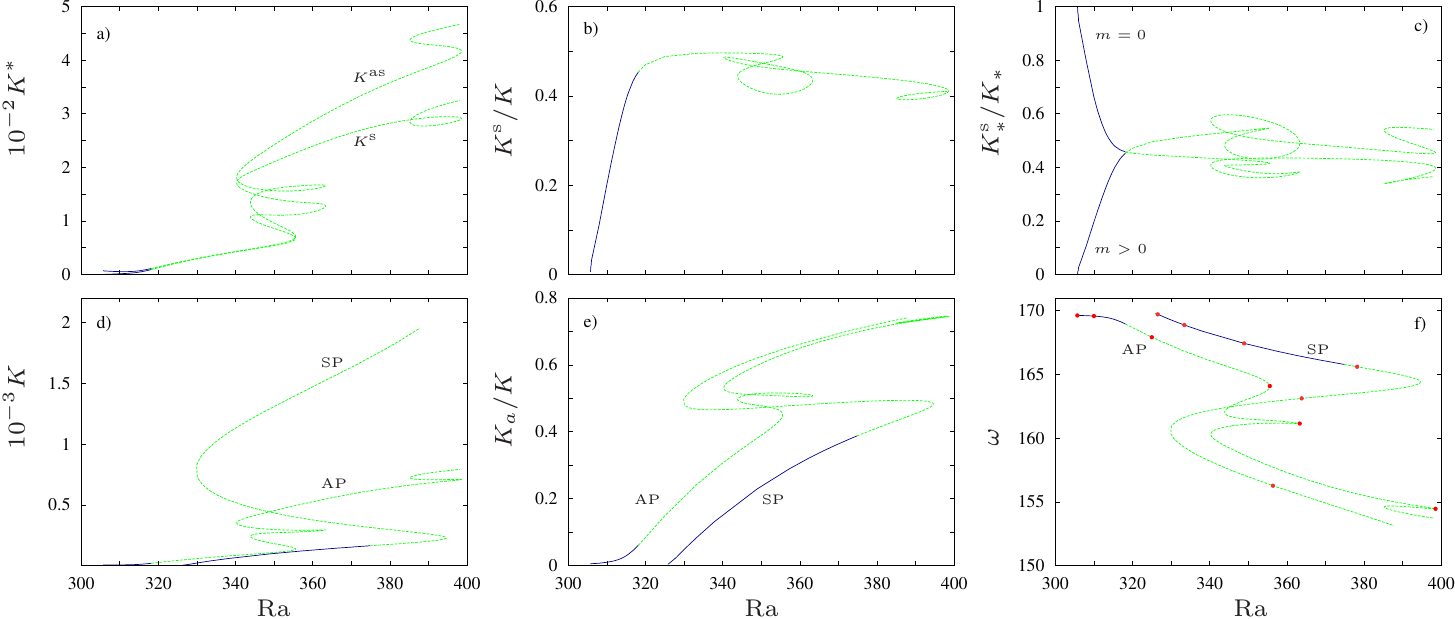}
\caption{Bifurcation diagrams of time and volume averaged quantities
  for varying $\Ray$. Panels (a), (b), and (c) are for the AP RW
  branch and (d), (e), and (f) also include the SP RW branch. (a)
  Kinetic energy density of the equatorially symmetric component of
  the flow $K^{\text{s}}$ and of the equatorially antisymmetric part
  $K^{{\text{as}}}=K-K^{\text{s}}$.  (b) Ratio of the equatorially
  symmetric kinetic energy density over total kinetic energy density
  $K^{\text{s}}/K$. (c) Ratio of the equatorially symmetric kinetic
  energy density of the axisymmetric ($m=0$) flow over total
  axisymmetric kinetic energy density $K_a^{\text{s}}/K_a$ and the
  same ratio $K_{na}^{\text{s}}/K_{na}$ for the nonaxisymmetric
  ($m>0$) flow. (d) Total kinetic energy density $K$.  (e) Ratios of
  the axisymmetric kinetic energy over the total kinetic energy
  density $K_a/K$. (f) Rotation frequency. Solid/dashed lines mean
  stable/unstable RW. The points (circles) of panel (f) correspond to
  RW shown in Fig.~\ref{fig:cplt_vphi_con} and
  Fig.~\ref{fig:cplt_vphi_mer}.}
\label{fig:bif_diagr}   
\end{figure*}

According to the linear study~\cite{GCW18}, at $\eta=0.9$,
$\Pra=3\times 10^{-3}$ and $\Ta=10^7$ the basic conductive state is
unstable to nonaxisymmetric perturbations with $m_d=19$ azimuthal wave
number and equatorial antisymmetry located near the poles. The
critical Rayleigh number is $\Ray_c=3.056\times 10^2$ and the critical
frequency is $\omega_c=-3225$ (see Fig. 5 of~\cite{GCW18}). Because
the azimuthal symmetry of the basic state is broken~\cite{EZK92} a
Hopf bifurcation gives rise to a rotating wave (also called travelling
wave) drifting in the azimuthal direction with rotation frequency
$\omega=-\omega_c/m_d$. We recall that any azimuthally averaged
property of a rotating wave will be constant, or in other words, a
rotating wave is a steady solution in the system of reference rotating
with frequency $\omega$. Notice that at the bifurcation the equatorial
symmetry of the basic state is also broken and the branch of rotating
waves is then equatorially asymmetric. From now on we use the term AP
RW to denote an equatorially asymmetric polar rotating wave. In
addition to the branch of AP RW that bifurcates first from the basic
state, we also trace a branch of equatorially symmetric rotating waves
(SP RW from now on) associated with the second preferred $m_d=19$
linear mode (with $\Ray_c=3.25\times 10^2$ and $\omega_c=-3169$).

The continuation method described in Sec.~\ref{sec:co_meth} is used to
obtain the branch of AP RW as a function of $\Ray$. To start the
continuation process the eigenfunction provided by the linear
stability analysis~\cite{GCW18} is used as initial guess. We consider
$N_r=30$ radial collocation points and $L_{\text{max}}=114$ spherical
harmonics truncation parameter with time steps around $\Delta
t=5\times 10^{-6}$. An $m_d=19$-fold azimuthal symmetry is assumed to
obtain a RW to speed-up the computations in the continuation process,
but all of the spherical harmonics should be considered to study the
stability, since the symmetry of the RW's dominant eigenfunction is
unknown. The spatial resolution is increased up to $N_r=50,~L=190$ to
check the computations, and a comparison of several outputs of a RW at
$\Ray=3.1\times 10^2$ is summarized in
Table~\ref{table:mesh_study}. The values of the rotation frequency
$\omega$, volume-averaged kinetic energy $K=\frac{1}{2}\langle
|\mathbf{v}|^2 \rangle_{\mathcal V}$ , and the value of the dominant
complex Floquet multiplier (from the stability analysis, see
Sec.~\ref{sec:co_meth}), remain almost unchanged by increasing the
resolution. As an additional numerical test, we obtain a stable AP RW
by means of DNS with $N_r=30$ and $L_{\text{max}}=84$ which is still
well resolved, giving rise to less than a $5\%$ difference in the
kinetic energy density and rotation frequency with respect to the AP
RW computed with continuation methods. The results for the stability
analysis of the waves are validated further by filtering (i.~e.
time-stepping several rotation periods) the initial guess for the
eigenvalue solver to avoid spurious eigenvalues (see~\cite{SGN16} for
a discussion). The number of dominant eigenvalues requested in the
ARPACK package is usually 16, but we have increased the value up to 40
to check the results. With these considerations the present study
constitutes a very large application of continuation methods in fluid
dynamics~\cite{DWCDDEGHLSPSSTT14}: it investigates a large number of
RW in a regime with a large number of degrees of freedom $n=1150517$
($n=5362609$, when $N_r=50,~L=190$) and reveals immense complexity not
explained by direct numerical simulations (DNS) but nonetheless vital
for understanding real systems. Although with DNS the different mode
contributions of the flow can be described, continuation techniques
should be used to understand their origin. According
to~\cite{RuTa71,Eck81,Hof_et_al04,KUL12} turbulent flows are described
in terms of unstable solutions (periodic or quasiperiodic) which can
not be captured by DNS.

\begin{figure*}
\includegraphics[scale=1.15]{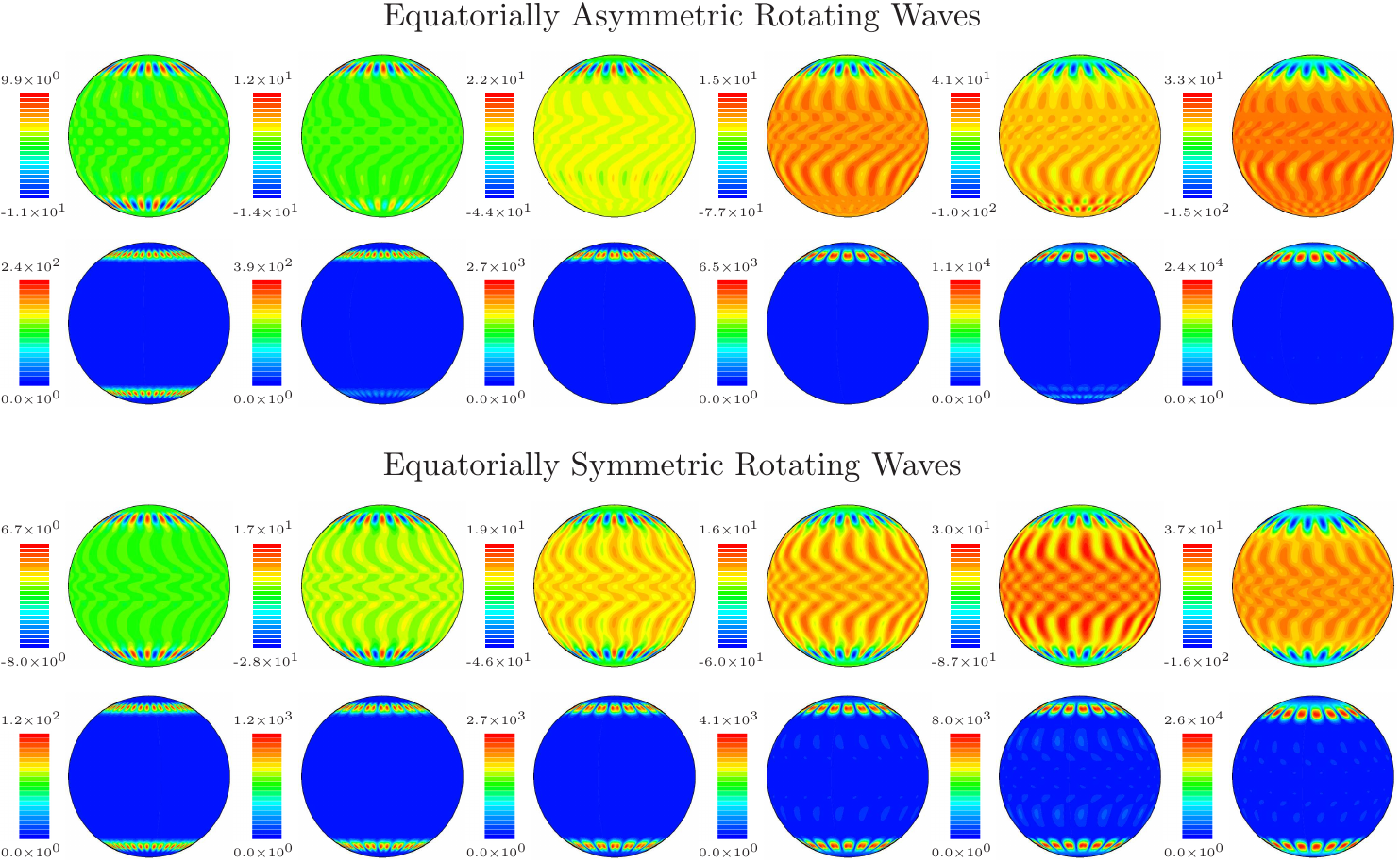}
\caption{1st-2nd rows: Contour plots of $m=19$ RW on the AP branch at
  $\Ray=3.057\times 10^2, 3.1\times 10^2, 3.25\times 10^2, 3.56\times
  10^2, 3.63\times 10^2, 3.99\times 10^2$ (from left to right). 1st
  row: Spherical sections of $v_{\phi}$ at $r_o$. 2nd row: Spherical
  sections of $\ve^2/2$ at $r_o$. 3rd-4th rows: As previous rows but
  for RW on the SP branch at $\Ray=3.27\times 10^2, 3.33\times 10^2,
  3.49\times 10^2, 3.78\times 10^2, 3.64\times 10^2, 3.56\times 10^2$
  (from left to right). All waves are marked with a circle in
  Fig.~\ref{fig:bif_diagr}(f).}
\label{fig:cplt_vphi_con}  
\end{figure*}

\begin{figure*}
\includegraphics[scale=1.15]{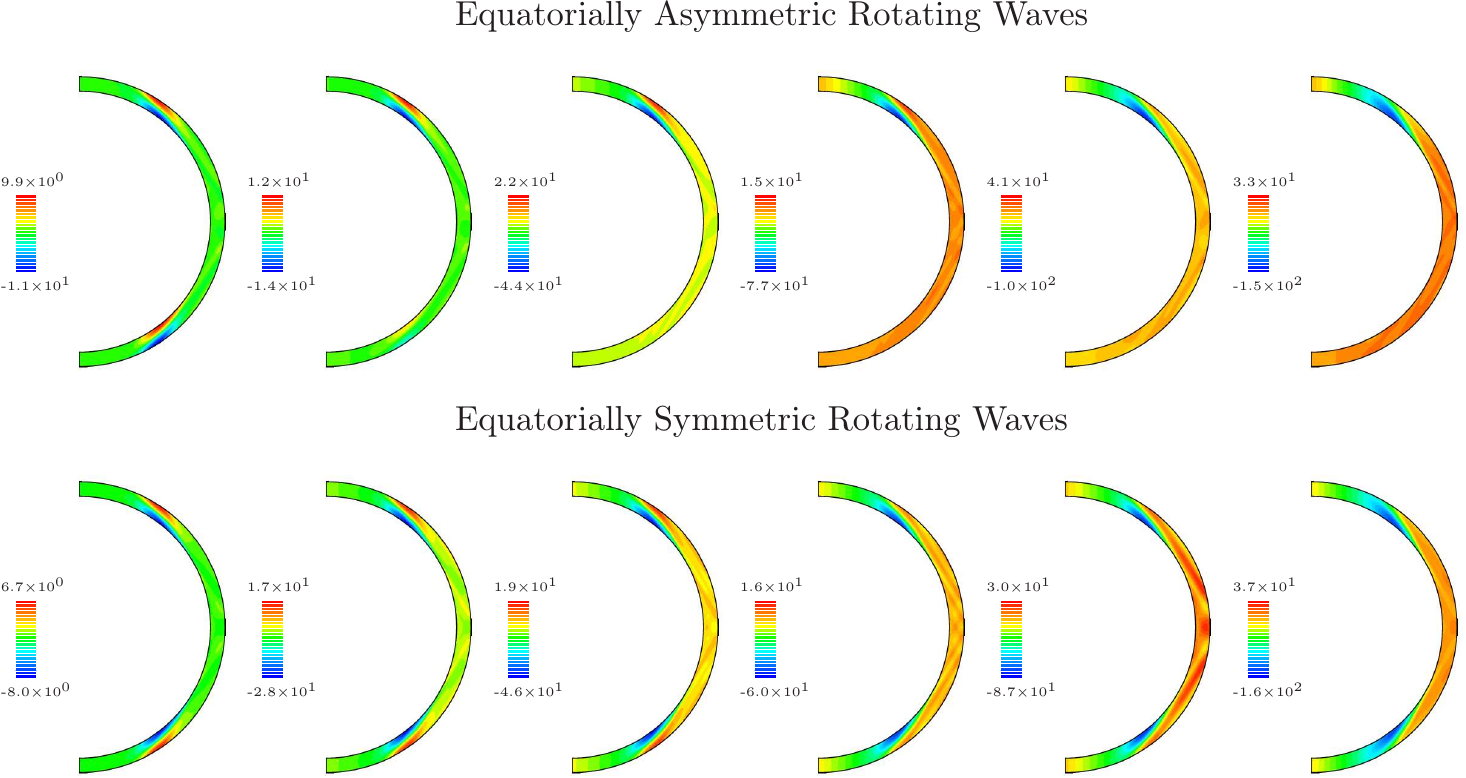}
\caption{Meridional sections of $v_{\phi}$ through a relative
  maximum. 1st row: Contour plots of $m=19$ RW on the AP branch at
  $\Ray=3.057\times 10^2, 3.1\times 10^2, 3.25\times 10^2, 3.56\times
  10^2, 3.63\times 10^2, 3.99\times 10^2$ (from left to right). 2nd
  row: As previous row but for RW on the SP branch at $\Ray=3.27\times
  10^2, 3.33\times 10^2, 3.49\times 10^2, 3.78\times 10^2, 3.64\times
  10^2, 3.56\times 10^2$ (from left to right). All waves are marked
  with a circle in Fig.~\ref{fig:bif_diagr}(f).}
\label{fig:cplt_vphi_mer}  
\end{figure*}

\begin{figure*}[t!]
\includegraphics[scale=1.15]{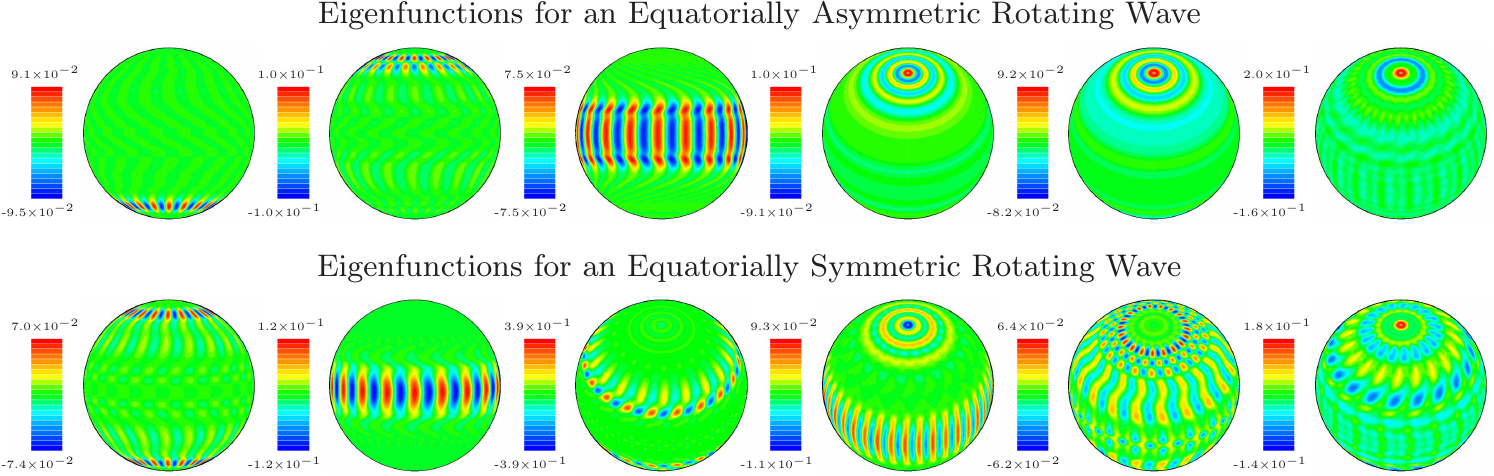}
\caption{First row: Contour plots of 1st, 4th, 6th, 7th, 9th and 13th
  (from left to right) dominant eigenfunctions on the AP RW branch at
  $\Ray=3.2\times 10^2$. Spherical section of $v_{\phi}$ at
  $r_o$. Second row: As first row but for the 1st, 2nd, 7th, 8th, 9th
  and 10th dominant eigenfunctions on the SP RW branch at
  $\Ray=3.78\times 10^2$.}
\label{fig:cplt_vphi_vep}  
\end{figure*}

Figure~\ref{fig:bif_diagr}(a,b,c) displays bifurcation diagrams (of
time and volume averaged data) for the AP branch of RW. The stability
region in each branch of RW is marked with a solid line. The branches
are started slightly above the point where they bifurcate from the
conductive state (best seen in Fig.~\ref{fig:bif_diagr}(a)). This is
because the convergence of the GMRES in the Newton iteration degrades
when RW are close to the onset (see~\cite{SaNe15} for an illustrative
example).  The kinetic energy density contained in the equatorially
symmetric flow $K^{\text{S}}$ increases from zero very sharply after
the onset.  The equatorially antisymmetric kinetic energy density
$K^{\text{AS}}=K-K^{\text{S}}$ is larger in all of the branch (see
Fig.~\ref{fig:bif_diagr}(a,b)) meaning that the flow is strongly
equatorially asymmetric very close to the onset. The axisymmetric
$m=0$ component of the flow abruptly loses its equatorially symmetry,
while by contrast the nonaxisymmetric ($m>0$) component becomes more
equatorially symmetric, see Fig.~\ref{fig:bif_diagr}(c). This is an
unexpected result, a strong equatorially symmetric $m=0$ flow
component was found in previous studies (see for
instance~\cite{HAW05,GWA13,GOD17}) mainly at larger $\Pra$ but also at
the low $\Pr$ regime~\cite{KSVC17}. Note that
$K_{a}^{\text{s}}/K_{a}\sim K_{na}^{\text{s}}/K_{na}\in (0.3,0.6)$.  A
strongly supercritical $\Ray$ regime is needed to obtain flows with
$K_{na}^{\text{s}}/K_{na}\in (0.3,0.6)$ at larger $\Pra$ in thicker
shells (see~\cite{GOD17} covering a large dynamo and hydrodynamical
database), contrasting to the present study near the onset. In
Fig.~\ref{fig:bif_diagr}(d,e,f) the SP RW branch is included. Its
kinetic energy density is significantly larger than the corresponding
value on the AP branch (see Fig.~\ref{fig:bif_diagr}(d)) because
convection is nearly absent in the south hemisphere as a result of the
strong equatorial asymmetry. For both AP and SP RW, the axisymmetric
component of the flow (Fig.~\ref{fig:bif_diagr}(e)) rises strongly,
reaching almost $80\%$ at the largest $\Ray$ explored. Strong zonal
flows are characteristic of gas giants~\cite{Chr02}. In addition, the
Rossby number $\Ros=\Ta^{-1/2}\sqrt{2K}<2\times 10^{-2}$ is small
(from Fig.~\ref{fig:bif_diagr}(d)) on both branches indicating the
importance of the Coriolis force compared to inertial forces, as
occurs in Jupiter's zonal flows~\cite{ScLi09}. Similar rotation
frequencies (time scales) are obtained for all of the waves
(Fig.~\ref{fig:bif_diagr}(f)), but this is not surprising as critical
frequencies at the onset are very similar.

\begin{figure*}
\includegraphics[scale=1.15]{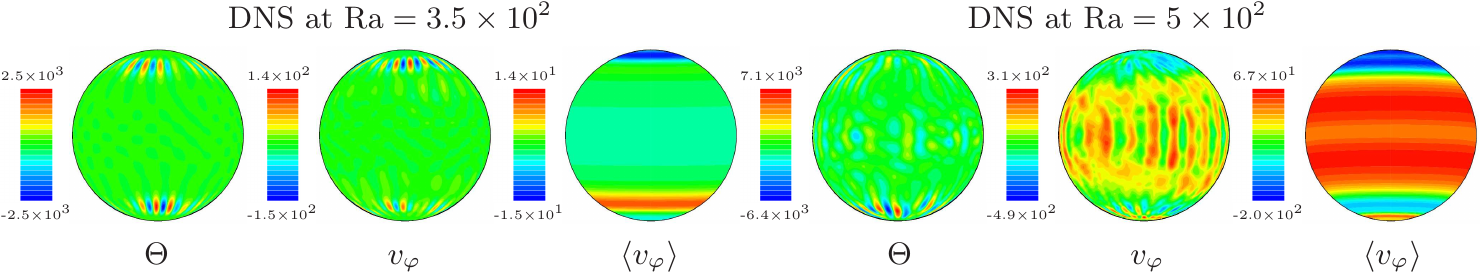}
\caption{Instantaneous contour plots of DNS at $\Ray=3.5\times 10^2$
  (left group of three plots) and $\Ray=5\times 10^2$ (right
  group). For each group the spherical sections (from left to right)
  are of $\Theta$ (at $r\approx r_i+0.5d$), $v_{\varphi}$ and $\langle
  v_{\varphi}\rangle$ (at $r=r_o$), respectively.}
\label{fig:cp_dns}   
\end{figure*}

The stability analysis, summarised in Sec.~\ref{sec:co_meth}, allows
us to obtain bifurcation points on the branches. Rather than computing
them accurately as in~\cite{SGN13,GNS16} by inverse interpolation
using several points on the branches, we roughly approximate the
bifurcation point to be at the point where $|\lambda|>1$
(resp. $|\lambda|<1$) is satisfied for the first time. This choice
makes sense, as the difference on the parameters between the previous
stable (resp. unstable) RW is small and $|\lambda|\approx1$. The AP RW
branch emerges from the preferred linear mode via supercritical Hopf
bifurcation at $\Ray_c=3.056\times 10^2$ and is thus stable according
to bifurcation theory~\cite{CrKn91,Kuz98}. The first bifurcation on
the AP RW branch is also of Hopf type at roughly $\Ray=3.18\times
10^2$. Surprisingly, the SP RW branch has a wider stable region, from
$\Ray=3.26\times 10^2$ to $\Ray=3.75\times 10^2$. The SP RW branch
bifurcates unstable as it comes from the 2nd preferred linear mode
with $m_d=19$-fold azimuthal symmetry although this is not noticeable
in Fig.~\ref{fig:bif_diagr} because SP RW restabilize very close to
the onset. This numerically proves that nonpreferred linear modes
(with the same $m_d$-fold azimuthal symmetry as the prefered mode)
contribute to stable flows as is argued in the
literature~\cite{GSN08,LaAu11}. It is an important result as it is
usual to compute only the first preferred linear mode (with fixed
$m_d$) to determine the onset of
convection~\cite{AHA04,DSJJC04,ViSc15,KSVC17}.

A branch of stable modulated rotating waves (MRW), which are
quasiperiodic flows (see~\cite{Ran82,CoMa92} for a theoretical
description), emerges when RW lose their stability via supercritical
Hopf bifurcations at $\Ray=3.18\times 10^2$ (AP RW branch) and
$\Ray=3.75\times 10^2$ (SP RW branch). In addition, stable or unstable
MRW could indeed be present near $\Ray=3.26\times 10^2$, i.~e. where
RW on the SP branch become stable, depending on whether the
bifurcation is subcritical or supercritical. The complex Floquet
multipliers $\lambda_k$ are quite clustered near the unit circle, for
instance at $\Ray=3.73\times 10^2$ on the SP RW branch
$0.88<|\lambda_k|< 1$ for the $k=1,..,10$ first dominant
multipliers. This leads to a large number of Hopf bifurcations, giving
rise to oscillatory flows which can be obtained with continuation
methods following~\cite{GNS16}. Because of the large number of
bifurcation points and the clustering of the eigenvalues, very long
initial transients ($O(10^2)$ diffusion time units) are expected if
DNS are used. Moreover, because RW with azimuthal wave number close to
$m=19$ are also expected to be stable near the onset
(see~\cite{GNS16,GCW18}) multistability regions of several RW and MRW
involving different azimuthal as well as equatorial symmetries could
be found, giving rise to very rich nonlinear periodic, quasiperiodic
and even chaotic dynamics very close to the onset ($\Ray/\Ray_c<1.7$).

The azimuthal velocity $v_{\varphi}$ patterns on the outer surface
along the AP and SP branches of RW are shown in the first and second
rows of Fig.~\ref{fig:cplt_vphi_con}.  Convection in the southern
hemisphere is progressively inhibited as $\Ray$ (i.~e. nonlinearity)
is increased from the onset on the AP branch. Around $\Ray\lesssim
3.2\times 10^2$ RW are stable, with most convection confined near the
north pole.  With increasing $\Ray$, positive $v_{\varphi}$ cells on
the south hemisphere increase their magnitude but remain significantly
weaker than the negative $v_{\varphi}$ cells surrounding the north
pole. Along all of the AP RW branch, the kinetic energy density is
concentrated near the north pole as well. This is a quite unexpected
result since it is widely assumed that finite amplitude convection
develops in both hemispheres (see~\cite{Bus70a,ABW97,HAW05,KSVC17}
among many others).  In the case of the SP RW, convection develops on
both hemispheres as is common and, like in the AP RW branch,
$v_{\varphi}$ is positive on a wide equatorial belt and negative near
the poles, the latter being stronger. Characteristic patterns of
inertial waves studied in~\cite{RiVa97} can also be identified in both
classes of polar waves, best shown in the meridional sections of
$v_{\varphi}$ in the 4th column (from left to right) of
Fig.~\ref{fig:cplt_vphi_mer}. These are structures elongated in the
colatitudinal direction, reflecting in both boundaries and connecting
the flow within the tangent cylinder from north to south
latitudes. The connection of the $v_{\varphi}$ vortices from north to
south can be best identified on the spherical sections of the 4th
column (from left to right) in Fig.~\ref{fig:cplt_vphi_con}.

In Fig.~\ref{fig:cplt_vphi_vep} the same contour plots as
Fig.~\ref{fig:cplt_vphi_con} are shown for different dominant
eigenfunctions at $\Ray=3.2\times 10^2$ on the AP RW branch (1st row)
and at $\Ray=3.78\times 10^2$ on the SP RW branch (2nd row). These
$\Ray$ are close to the bifurcations to stable MRW. In both branches
the patterns of the first dominant eigenfunction are quite similar to
that of the RW and thus similar patterns are expected for the MRW. In
contrast, other eigenfunctions behave quite differently, reflecting
the multimodal character of low-$\Pr$ number
convection~\cite{ABGHV18}. The eigenfunction patterns can be
equatorially symmetric of equatorial type, or in the form of strongly
axisymmetric belts surrounding the poles (see right three plots). We
note the good agreement of the polar belts of the eigenfunction with
those obtained in the experimental study of~\cite{CAFL17} in the
context of zonal flows in giant planets. In addition, this
eigenfunction has convection confined within the $80^\circ$ latitude
circle, towards the poles, as observed on the surfaces of
Jupiter~\cite{Adr_et_al18} or Saturn~\cite{SBDEI17}.

\subsection{Chaotic flows from DNS}

\begin{figure*}[t!]
  \includegraphics[scale=1.15]{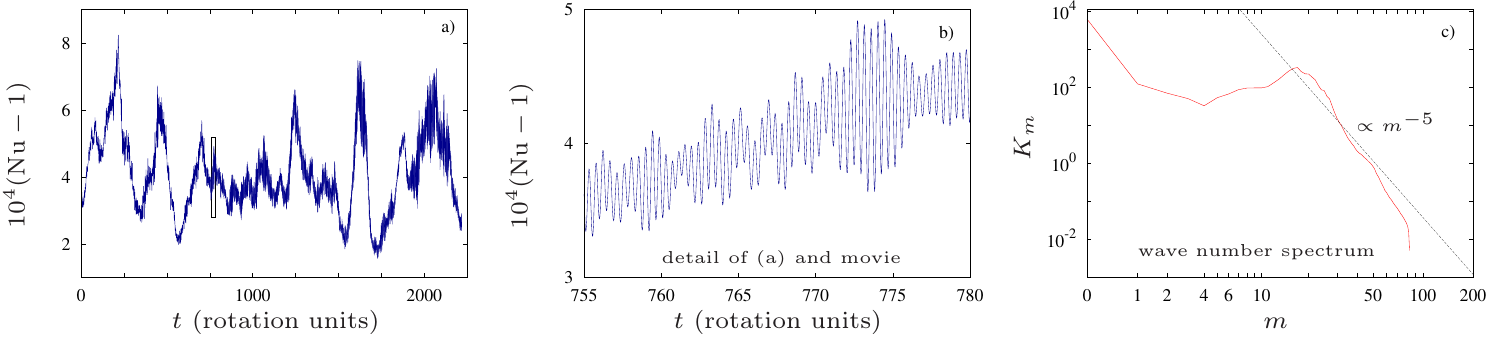}
  \caption{(a) Time series of convective heat transfer at the outer
    surface $\Nus-1$ and (b) a detail of (a) including the time span
    of the supplementary movies~\cite{Sup00}. (c) The kinetic energy spectrum $K_m$
    versus the azimuthal wave number $m$ and the
    theoretical~\cite{Rhi75} Rhines' scaling (dashed line).}
\label{fig:ts_dns}      
\end{figure*}

\begin{table*}[t!] 
  \begin{center}
    \begin{tabular}{lcccccccccc}
\vspace{0.1cm}           
& $\Ray$           & $\Ray/\Ray_c$ & $\Ree$ & $\Ros$   & $\Ros_c$ & $\Pec$  & $K_a/K$& $\mathcal{F}_C$  & $\mathcal{F}_I$ & $\mathcal{F}_V$     \\
\hline\\[-8.pt]
& $3.5\times 10^2$~ & $1.15$~      & $37$~  & $0.012$~ & $0.34$~  & $1.1$~  & $0.04$ & $1.2\times 10^5$~ & $1.8\times 10^4$~ & $9.5\times 10^2$~     \\
& $5\times 10^2$   & $1.64$        & $142$~ & $0.045$  & $0.41$   & $4.3$   & $0.64$ & $6.3\times 10^5$ & $1.2\times 10^5$ & $3.6\times 10^3$     \\
\hline
  \end{tabular}
    \caption{Time and volume averaged properties for DNS at
      $\Pra=3\times 10^{-3}$, $\Ta=10^7$ and $\eta=0.9$, with $\Ray$
      close to the onset of convection. The Reynolds number $\Ree$,
      the Rossby number $\Ros$, the convective Rossby number $\Ros_c$,
      the Peclet number $\Pec$, and the ratio of axisymmetric ($m=0$)
      over total rms kinetic energies $K_a/K$, and the rms of forces
      integrals $\mathcal{F}_I$ (inertial), $\mathcal{F}_C$ (Coriolis)
      and $\mathcal{F}_V$ (viscous) are tabulated.}
  \label{tab_mean}
  \end{center}
\end{table*}

Like the very recent low-Prandtl number studies~\cite{HoSc17,ABGHV18},
we find very rich quasiperiodic and even chaotic dynamics very close
to the onset ($\Ray/\Ray_c<1.7$) involving several basic modes of
convection and time scales. Two examples of these strongly oscillatory
DNS, obtained from an AP wave initial condition, are shown in
Fig.~\ref{fig:cp_dns} at $\Ray=3.5\times 10^{2}$ and $\Ray=5\times
10^{2}$. The left three plots of Fig.~\ref{fig:cp_dns} are
instantaneous contour plots of the temperature perturbation $\Theta$,
the azimuthal velocity $v_{\varphi}$, and the azimuthally-averaged
azimuthal velocity $\langle v_{\varphi}\rangle$, on a spherical slice
at $r\approx r_i+0.5d$, and $r=r_o$, respectively, for the DNS at
$\Ray=3.5\times 10^{2}$. The flow is strongly equatorially asymmetric
and located in polar regions, recalling an AP RW. In contrast,
convection at larger $\Ray=5\times 10^2$ (right group of three plots)
also develops in the equatorial region taking the form of vertical
columns. The latter resemble the patterns of the equatorial
eigenfunction shown in Fig.~\ref{fig:cplt_vphi_vep}. Notice that
$\langle v_{\varphi}\rangle$ is positive in a wide band showing a
strong dimple (relative minimum) near the equator.

The pattern of these DNS seems to resemble a superposition of the RW
mode and the eigenfunctions shown in
Fig.~\ref{fig:cplt_vphi_con}. This is not surprising, as unstable RW
and MRW define the framework of the phase space and thus drive chaotic
and the turbulent
dynamics~\cite{Eck81,Hof_et_al04,KUL12,DWCDDEGHLSPSSTT14}. Clearly,
the flow at $\Ray=5\times 10^{2}$ is strongly axisymmetric and
multimodal~\cite{HoSc17,ABGHV18}. Figure~\ref{fig:ts_dns}(a) displays
the time series of the convective heat transport $\Nus-1$ at the outer
boundary in time rotation units $t\sqrt{Ta}/2\pi$ ($t$ is the
dimensionless diffusion time). This figure, and its detail in
Fig.~\ref{fig:ts_dns}(b), reveals the chaotic and strongly oscillatory
character of the flow involving time scales from less than 1, to
around $300$ planetary rotations. Figure~\ref{fig:ts_dns}(b) contains
the time span of the supplementary movie~\cite{Sup00}. The
time-averaged kinetic energy wave number spectrum is shown in
Fig.~\ref{fig:ts_dns}(c), together with the corresponding theoretical
Rhines' scaling~\cite{Rhi75} for the strongly axisymmetric flows
relevant to planetary atmospheres. This scaling has been confirmed
with a recent laboratory model, including viscous dissipation, of high
latitude jets on Jupiter's surface~\cite{CAFL17}. Regarding the weakly
supercritical $\Ray$, the spectrum has a peak at the most unstable
mode $m=19$ at the onset and roughly approximates the Rhines's scaling
for the larger wave numbers. The good qualitative agreement with the
experimental spectrum shown in Fig.3 of~\cite{CAFL17} is
noticeable. This provides supporting evidence for the turbulent
character of the DNS for the low $\Pr$ and thin shell. At higher $\Pr$
and thicker shells, corresponding to the linear stability region of
equatorial or spiralling modes, higher supercritical conditions must
be reached for the onset of turbulence~\cite{PlBu05,GSN14}.

\subsection{Equatorially asymmetric zonal winds}

\begin{figure*}
 \includegraphics[scale=2.2]{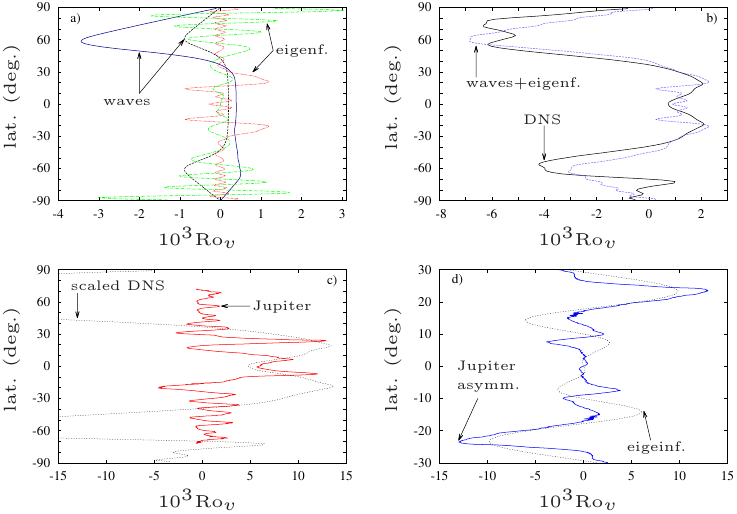}
 \caption{$\Ros_v=\langle v_{\varphi} \rangle/\Omega r_o$ versus
   latitude in degrees. (a) Asymmetric polar rotating wave
   $u_{\text{as}}$ at $\Ray=3.98\times 10^2$ (solid line), its 8th
   dominant eigenfunction (dotted-dashed line), symmetric polar
   rotating wave $u_{\text{s}}$ at $\Ray=3.78\times 10^{2}$ (dashed
   line) and its $u_{\text{2nd}}$ 2nd dominant eigenfunction (dotted
   line). The eigenfunctions are scaled for comparative purposes. (b)
   DNS at $\Ray=5\times 10^2$ (solid line) and the linear
   superposition $u_{\text{as}}+4u_{\text{s}}+1.5\times
   10^4u_{\text{2nd}}$ of the curves of $\Ros_v$ shown in (b) (dashed
   line).  (c) Hubble data for Jupiter~\cite{TWPSORACJMM17} (solid
   line) and DNS of (b) scaled by a factor of 6.3 (dotted line). (d)
   Hubble data for the antisymmetric Jupiter's zonal flow and the 8th
   dominant eigenfunction of the asymmetric polar rotating wave scaled
   by a factor of 6.5 (dotted line).}
\label{fig:lat_prof}   
\end{figure*}

In this section we qualitatively analyse the equatorial symmetry of
the zonal wind and describe some interesting properties of flows
studied in previous sections. The study of zonal wind is of
fundamental importance for the understanding of dynamics of giant
planet atmospheres atmospheres~\cite{HeAu07}, and in particular in the
case of Jupiter~\cite{HAW05}. The Taylor number of the present study,
$\Ta=10^7$, is moderate and thus far from the real
applications. However, it is interesting to investigate basic
properties of nonlinear flows arising from the onset of polar
convection, as this is common in thin shells $\eta=0.9$~\cite{GCW18},
which is believed to be a good geometry approximation for modelling
the convective envelope of Jupiter or Saturn~\cite{HeAu07}. In thin
shells, polar modes are linearly preferred also at $\Ta=10^{11}$ and
$\Pr=10^{-2}$, very close to $\Ta=10^{11}$ and $\Pr=10^{-1}$
of~\cite{HAW05}, which successfully explained the number of azimuthal
jets and their width in the Jovian atmosphere.

Flows in planetary atmospheres are turbulent and strongly
axisymmetric. Some basic parameters for their study are the Reynolds
number $\Ree=\sqrt{2K}$ as a measure of convection, the Rossby number
$\Ros=\Ta^{-1/2}\Ree$ to quantify the relevance of Coriolis
force~\cite{HAW05}, the convective Rossby number
$\Ros_c=(1-\eta)^{-1/2}\sqrt{\Ray/(\Ta\Pra)}$ (the factor $1-\eta$ is
due to our definition of $\Ray$) to determine the transition between
positive and negative equatorial jets~\cite{MMK15}, and the Peclet
number $\Pec=(1-\eta)^{-1}\Pra\sqrt{2K}$ used to define the boundary
($\Pec=10$) between weak and strong flows~\cite{KSVC17}. Flows in gas
giant atmospheres have small $\Ros$ (for instance $\Ros=0.01-0.04$
given in~\cite{HeAu07}) and $\Ros_c$ less than unity ($\Ros_c=0.22$
from the simulations of~\cite{HeAu07}) as their equatorial jets are
positive. In addition, $\Pec$ should be large as Jupiter's and
Saturn's atmospheric flows are vigorous. Table~\ref{tab_mean}
summarises these mean physical properties from the DNS at
$\Ray=3.5\times 10^2$ and $\Ray=5\times 10^2$. Both DNS have a
noticeable $\Pec>1$ and thus convection is getting stronger, although
they are close to the onset, as happened for similar $\Pr$ numbers in
a full sphere at large $\Ta$~\cite{KSVC17}, this reflects on a
relatively large $\Ree$. The Coriolis force is relevant as indicates
the small value of $\Ros$. For $\Ray=5\times 10^2$ the ratio of
axisymmetric to total kinetic energy density is relatively large
meaning the DNS is mostly axisymmetric and the equatorial zonal wind
belt is positive with $\Ros_c=0.41<1$, in reasonable agreement
with~\cite{MMK15} (at $\Pr=0.27$). Following~\cite{SKA12} the rms
force integrals $\mathcal{F}_I$ (inertial), $\mathcal{F}_C$ (Coriolis)
and $\mathcal{F}_V$ (viscous) are computed. The balance
$\mathcal{F}_C>\mathcal{F}_I\gg\mathcal{F}_V$, satisfied for the DNS,
is believed to operate in Jupiter's convective
atmosphere~\cite{ScLi09}. It is noticeable how the DNS properties detailed
above are in reasonable agreement with those of giant planets,
despite the fact that the parameter values are far from those
objects. Larger Taylor numbers and supercritical regimes, as
in~\cite{HAW05}, should be attained to study the nonlinear saturation
of polar modes, which could be of interest for planetary atmospheres
as the results of table~\ref{tab_mean} suggest.

\begin{figure}
  \includegraphics[scale=1.5]{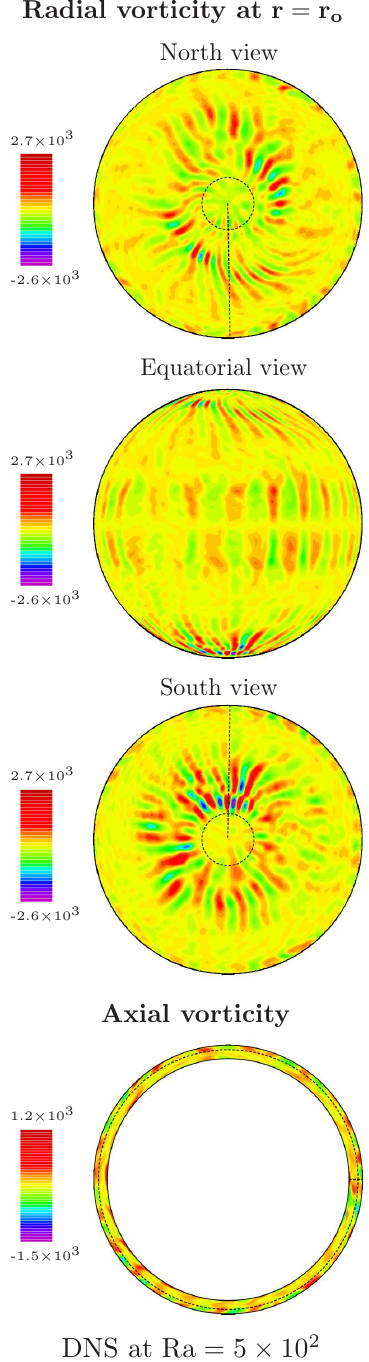}
  \caption{Radial vorticity at $r=r_o$ (viewed from the north pole,
    the equator and the south pole) and axial vorticity on the
    equatorial plane (from left to right).  Cyclonic (anticyclonic)
    radial vorticity is red (blue) on the northern hemisphere and blue
    (red) on the southern hemisphere. A supplementary movie~\cite{Sup00} for each
    panel is included.}
\label{fig:cp_vor}      
\end{figure}

Instantaneous latitude profiles of $\Ros_v=\langle
v_{\varphi}\rangle/\Omega r_o$ at the outer surface for an AP and SP
RW, and two examples of eigenfunctions (properly scaled), are shown in
Fig.~\ref{fig:lat_prof}(a) for a qualitative comparison.  Both classes
of RW have off-equatorial (around $60^\circ$) negative and positive
equatorial zonal bands, but the profile for the AP RW is strongly
asymmetric, being positive around $-60^\circ$. An eigenfunction of
equatorial type gives rise to a large number of bands, with two strong
relative minima and maxima for latitudes in
$(-20^\circ,20^\circ)$. All of these characteristics can be identified
in the equatorially asymmetric zonal wind profile of the chaotic DNS
at $\Ray=5\times 10^2$, see Fig.~\ref{fig:lat_prof}(b). A suitable
superposition of the profiles shown in Fig.~\ref{fig:lat_prof}(a) at
lower $\Ray$ reasonably resemble the profile of the chaotic DNS at
$\Ray=5\times 10^2$, both showing a strong dimple (decrease of
positive velocity) at the equator as observed in
Fig~\ref{fig:cp_dns}. This is reasonable as branches of RW described
in this study (and secondary MRW arising from unstable Floquet modes)
may extend up to $\Ray=5\times 10^2$.  The magnitude of the zonal wind
clearly differs from that measured in the Jovian atmosphere, we note
however that the decrease at the equator is larger than $50\%$ and
qualitatively comparable to the decrease observed in the Jovian
atmosphere (see Fig.~\ref{fig:lat_prof}(c) revealed by Hubble Space
Telescope observations ~\cite{TWPSORACJMM17}). In anelastic
models~\cite{GWA13}, it was argued that the dimple was related to the
existence of a change in the dynamic behaviour with two different flow
dynamics within the shell. In our case the situation seems to be
similar, the flow being more influenced by Coriolis forces in the
interior of the shell. Figure~\ref{fig:cp_vor} (right panel) contains
an equatorial section of the axial vorticity, as in~\cite{GWA13} for
the anelastic models, and displays the situation. A dashed line is
drawn to mark the boundary between two different dynamical
behaviours. Finally Fig.~\ref{fig:lat_prof}(d) shows the Hubble Space
Telescope data for the asymmetric component of the flow, compared to a
selected (properly scaled) eigenfunction of an AP RW. It is noticeable
the qualitative similarities between both profiles which invites to
further research of asymmetric modes such as described in the present
study, but at larger $\Ta$, to see if they are relevant for
understanding the equatorial asymmetry of Jovian
atmosphere~\cite{TWPSORACJMM17}

Figure~\ref{fig:cp_vor} displays instantaneous contour plots of the
radial vorticity, at the outer surface in different views, of the
chaotic DNS at $\Ray=5\times 10^2$. It shows the existence of large
coherent vortices in the equatorial belt, but also at very high
latitudes, even within the $80^\circ$ circle (dashed line) towards the
poles. In the Jovian atmosphere large vortices develop in the
equatorial region, as found numerically in~\cite{HGW15} at higher
$\Ta=10^{11}$ with a stratified model in thin shells. Large coherent
cyclonic vortices, surrounding the poles, have recently been observed
in Jupiter's atmosphere~\cite{Adr_et_al18}. Our simulated vortices at
high latitudes are quite elongated in the meridional direction and
obtained at moderate $\Ta$ and thus not representative of the real
situation. In addition, the ratio between the number of cyclonic and
anticyclonic vortices is not as large as for the Jovian atmosphere.
However our results provide evidence that cyclonic coherent vortices
might be obtained at high latitudes as well at larger $\Ta$, since
they are strongly related to AP or SP linear modes which are preferred
at relevant $\Ta>10^{10}$ as well. As commented before, this makes the
regime of polar modes interesting and further challenging simulations
are thus required to see if these type of convection is relevant for
the basic understanding of the appearance of large scale coherent
structures at high latitudes in spherical shell convection models.

\section{Summary}
\label{sec:sum}

A numerical study of thermal convection in rotating spherical shells
is presented. The parameter values $\eta=0.9$, $\Ta=10^7$,
$\Pr=3\times 10^{-3}$ are selected to study the unexplored regime when
convection onsets from equatorially antisymmetric and nonaxisymmetric
polar modes~\cite{GSN08,GCW18}. This contrasts to the regimes studied
widely for many years~\cite{Bus70a,SiBu03,HAW05,KSVC17} (among many
others) in which convection onsets from an equatorially symmetric and
nonaxisymmetric perturbation, either equatorially
attached~\cite{Zha93,PlBu05} or spiralling~\cite{Zha92}.

As with most studies in the field~\cite{SiBu03,HAW05,KSVC17} the
numerical study is based on using DNS to obtain chaotic
flows. However, there exist very few studies~\cite{SGN13,FTGS15} in
which continuation techniques and the stability analysis of periodic
orbits (Floquet theory) are employed to track the curves and determine
the regions of stability of rotating waves (RW) bifurcated from the
conductive state. Tracking these unstable branches is of fundamental
importance for a deep understanding of the origin of chaotic and
turbulent flows~\cite{RuTa71,Eck81,Hof_et_al04,KUL12}.

We have obtained stable equatorially asymmetric (AP) as well as
symmetric polar (SP) RW, the latter associated with the 2nd dominant
linear mode with $m=19$-fold azimuthal symmetry. In this case,
non-preferred linear modes - with the same azimuthal symmetry as the
preferred mode - can give rise to stable flows and should be computed
in linear studies. This is relevant, as most studies have relied on
the computation of the first preferred linear
mode~\cite{AHA04,DSJJC04,ViSc15,KSVC17} and this mode is usually used
to initialize DNS with parameters close to the onset.

The patterns of the AP/SP RW are steadily rotating in the azimuthal
direction, with the flow developing a strong axisymmetric component and
confined at high latitudes. Surprisingly, convection is almost
hemispherical in the case of AP RW, in contrast to what has been found in
previous studies. In addition, in our simulations RW (i.~e. steadily
azimuthally drifting periodic flows) are obtained at the low $\Pr$
regime and $\Ta=10^{7}$: it is not clear if they still are
present when $\Ta$ is increased; see~\cite{KSVC17} for a full sphere
or the rotating convection experiments of~\cite{ABGHV18}.

By means of DNS at $\Ray/\Ray_c<1.7$, oscillatory chaotic flows are
obtained, in agreement with low-Prandtl number
studies~\cite{KSVC17,HoSc17,ABGHV18}. Their dynamics are strongly
influenced by unstable RW and quasiperiodic flows (modulated rotating
waves), related to the eigenfunctions of the RW, as our numerical
results suggest. Computation of stable/unstable waves then provides a
useful tool to understand the dynamics of turbulent
flows~\cite{RuTa71,Hof_et_al04}.

In addition, these oscillatory flows reveal physical regimes which
share qualitative characteristics with those occurring in the
Jovian atmosphere. They have strong zonal and equatorially asymmetric
components, including the presence of polar coherent
vortices~\cite{Adr_et_al18} and the characteristic
dimple~\cite{TWPSORACJMM17}.  Although the value $\Ta=10^{7}$ of our
models is still small, the force balance of the simulations is in
concordance with that believed to operate in the planetary atmospheres
and the flow physical properties such as the Rossby
number~\cite{HAW05} or the convective Rossby number~\cite{MMK15} are
not so far from their estimated values.  Our study indicates the
  need for further research at more relevant $\Ta>10^{10}$, as polar
  convection is linearly preferred also in this regime.

Our results may also be relevant in explaining the coherent structures
without equatorial symmetry, vortices, observed on Saturn's polar
surface in late 2012 by the Cassini spacecraft~\cite{SBDEI17}. The
Polar convection is known to grow in strongly nonlinear
regimes~\cite{Chr02}, but we have shown the existence of polar flows
at very low Rayleigh numbers. Our results may also help for the
understanding planetary core dynamos, as AP modes may be preferred at
the onset with nonslip conditions~\cite{GSN08}. According
to~\cite{GOD17}, equatorially asymmetric flows favour the appearance
of multipolar magnetic fields, which have been shown to be
characteristic of thin shells~\cite{SPD12,OrDo14b}. Understanding the
equatorial symmetry breaking of the flow and its associated
transitions is then of importance in planetary dynamo numerical
models.

Because very low $\Pra$ and large $\eta$ are common in stellar
convective zones~\cite{GCW18} our results are also of importance for
the astrophysical community. A transition between positive and negative
zonal flow profiles was found in~\cite{MMK15} in the context of
stellar magnetohydrodynamic flows, such as those appearing in the Sun.
In the case of accreting neutron star oceans,
convection~\cite{ZMNAB15}, zonal flows~\cite{Spitkovsky02}, pattern
formation and coherent structures~\cite{Wat12}, are key issues for a
deeper understanding of thermonuclear X-ray bursting phenomena.

\medskip
\section{Acknowledgements}

F. G. was supported by a postdoctoral fellowship of the Alexander von
Humboldt Foundation. The authors acknowledge support from ERC Starting
Grant No. 639217 CSINEUTRONSTAR (PI Watts). This work was sponsored by
NWO Exact and Natural Sciences for the use of supercomputer facilities
with the support of SURF Cooperative, Cartesius pilot project
16320-2018. The authors wish to thank F. Stefani, M. Net, and
J. S\'anchez for useful discussions.

\end{document}